\documentclass[prl,twocolumn,superscriptaddress,showpacs,epsf]{revtex4}

\usepackage{epsfig}
\usepackage{color}

\usepackage{graphicx}
\usepackage{latexsym}
\usepackage{amsmath}
\usepackage{amssymb}
\usepackage{amsfonts}

\begin{document}

\title{Planar superconductor/ferromagnet hybrids: Anisotropy of resistivity induced by magnetic templates}

\author{A.Yu. Aladyshkin}
\affiliation{INPAC -- Institute for Nanoscale Physics and
Chemistry, K.U. Leuven, Celestijnenlaan 200D, B--3001 Leuven,
Belgium} \affiliation{Institute for Physics of Microstructures
RAS, 603950, Nizhny Novgorod, GSP-105, Russia}
\author{J. Fritzsche}
\affiliation{INPAC -- Institute for Nanoscale Physics and
Chemistry, K.U. Leuven, Celestijnenlaan 200D, B--3001 Leuven,
Belgium}
\author{V.V. Moshchalkov}
\affiliation{INPAC -- Institute for Nanoscale Physics and
Chemistry, K.U. Leuven, Celestijnenlaan 200D, B--3001 Leuven,
Belgium}

\date{\today}
\begin{abstract}
We investigated experimentally the transport properties of a
superconducting cross-shaped aluminium microbridge fabricated on
top of ferromagnetic BaFe$_{12}$O$_{19}$ single crystal. It was
demonstrated that a one-dimensional domain structure in the
ferromagnetic substrate can induce the formation of
superconducting channels above magnetic domains. This leads to a
giant anisotropy of resistivity of the superconducting bridge,
caused by the appearance of continuous paths of supercurrents
flowing along domain walls.
\end{abstract}

\pacs{74.78.-w 74.78.Fk 74.25.Dw}
\maketitle

Hybrid superconductor-ferromagnet (S/F) structures have attracted
considerable attention in connection with the possibility to
control thermodynamic and transport properties of the S/F hybrids
by manipulating the magnetic state of the ferromagnetic
constituents
(\cite{Buzdin-RMP-05,Lyuksyutov-AdvPhys-05,Velez-JMMM-08,Aladyshkin-SuST-09}
and references therein). Provided an insulating layer prevents the
diffusion of Cooper-pairs from the superconductor to the
ferromagnet, the exchange interaction between superconducting and
ferromagnetic parts can be effectively suppressed and the
interaction between both subsystems occurs via slowly decaying
stray magnetic fields. Nonuniform magnetic field, induced by the
ferromagnet, can modify the conditions for the appearance of
superconductivity due to the effect of a local field compensation,
resulting in the field-induced superconductivity
\cite{Lange-PRL-03} and an exotic dependence of the
superconducting critical temperature $T_c$ on an applied magnetic
field $H$ \cite{Aladyshkin-PRB-03,Yang-Nature-04,Gillijns-PRB-07}.
An increase of the width of the equilibrium magnetization loop
$M(H)$ of the S/F hybrids, compared with plain superconducting
films, can be interpreted as an enhanced ``magnetic" pinning of
vortices by various magnetic textures: periodic arrays of magnetic
dots \cite{Morgan-PRL-98,vanBael-PRB-99} or irregular magnetic
bubble domains \cite{Lange-APL-02}. The magnetostatic interaction
between the vortices and the ``built-in" magnetic field is also
known to lead to the unusual field dependence of the electrical
resistance $R(H)$ of the S/F hybrids at temperatures close to the
superconducting critical temperature $T_{c0}$
\cite{Lange-PRL-03,Martin-PRL-97,Jaccard-PRB-98,Hoffmann-PRB-00}.

Recently, electrical transport in S/F hybrids at low temperatures
was studied for the following planar structures: Nb/Co
\cite{Vodolazov-PRB-05}, Al/CoPd \cite{Morelle-APL-06},
NbSe$_2$/Py \cite{Vlasko-Vlasov-PRB-08a}, MoGe/Py
\cite{Belkin-APL-08,Belkin-PRB-08}, Pb/Py
\cite{Vlasko-Vlasov-PRB-08b}. Superposition of the bias current
and the supercurrent that is induced by hard ferromagnets may lead
to a remarkable change of the current ($I$) -- voltage ($V$)
characteristics of superconducting bridges
\cite{Vodolazov-PRB-05,Morelle-APL-06}, what can be interpreted as
a ``current" compensation effect \cite{Milosevic-PRL-05}. The
tunable alignment of magnetic domains in low-coercive
ferromagnetic films, using an in-plane oriented external field
$H$, makes it possible to introduce a guided vortex motion in a
desirable direction  -- along the domain walls
\cite{Vlasko-Vlasov-PRB-08a,Belkin-APL-08,Belkin-PRB-08,Vlasko-Vlasov-PRB-08b}.

    \begin{figure}[b!]
    \includegraphics[width=70mm]{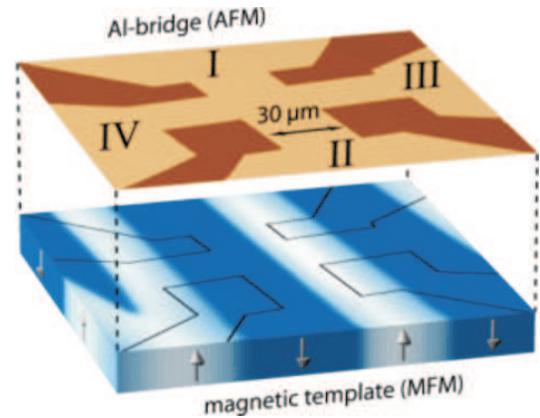}
    \caption{(color online) The planar S/F hybrid system under
    investigation. The top layer shows an atomic force microscopy
    image (AFM) of the cross-shaped Al bridge (lighter shades). The
    areas labelled I--IV were used as contact pads for transport
    measurements. The bottom layer shows a magnetic force microscopy
    image (MFM) of the ferromagnetic BaFe$_{12}$O$_{19}$ substrate.
    Light and dark regions correspond to the magnetic domains with
    $M_z>0$ and $M_z<0$. Note that the MFM image is
    vertically extended to illustrate the magnetic domains. Black
    solid lines depict the edges of the Al bridge.} \label{Fig-System}
    \end{figure}

In this Letter we are aiming at the investigation of the
anisotropy of the electrical transport properties of the S/F
hybrids, induced by \textit{a single} straight domain wall. We
measured both the magnetoresistance $R(H)$ and the $I-V$
dependencies of a superconducting bridge in two perpendicular
directions (i.e. along and across a domain wall in the
ferromagnetic substrate).

In order to study the anisotropy of the low-frequency transport
properties in a planar S/F hybrid, we prepared a bi-layered sample
consisting of a bulk ferromagnetic substrate and a thin-film Al
microbridge on top. The ferromagnetic and superconducting parts
were electrically isolated by a 5 nm SiO$_2$ buffer layer, so that
the interaction between these parts can be expected to be
exclusively electromagnetic. When cut along the proper
crystallographic direction, a ferromagnetic BaFe$_{12}$O$_{19}$
single crystal exhibits a one-dimensional (1D) stripe-type domain
structure with dominant in-plane magnetization and relatively
small out-of-plane component $M_z$ (the bottom image in
Fig.~\ref{Fig-System}). Measurements with a vibrating sample
magnetometer (VSM)  revealed that at low temperatures the
magnetization of the used crystal depends almost linearly on the
applied perpendicular magnetic field with the slope $dM/dH\simeq
3.2\cdot10^5$ (A/m)T$^{-1}$ and that it saturates at $H\simeq
1.7$~T. This means that external magnetic fields $|H|\le 80$~mT
can only be of minor influence on the domain structure, since the
variations of magnetic moment of the substrate are expected to be
about 4.5\% from the saturated magnetization
($5.5\cdot$10$^5$~A/m). The location of the domain walls and their
shape were determined by magnetic force microscopy (MFM), prior to
the preparation of the Al bridge. The expected amplitude of the
$z-$component of the nonuniform magnetic field, $B_0$, exceeds the
upper critical field $H_{c2}$ of such Al films even at low
temperatures (see below). The cross-shaped Al microbridge (50 nm
thick) was fabricated by means of e-beam lithography, molecular
beam epitaxy and lift-off etching (the top image in
Fig.~\ref{Fig-System}). The width $w$ of the "arms" of the
microbridge was equal to 30~$\mu$m and limited by the width of the
magnetic domains. Four contact pads, labelled in Fig.
\ref{Fig-System} as I--IV, were used for the injection of the dc
bias current $I$ and for the measurement of the voltage drop $V$
for two different cases: along the domain wall ($V_{\parallel}$)
using the electrodes I and II and across the domain walls
($V_{\perp}$) using the electrodes III and IV. This symmetrical
form of the superconducting element was intentionally chosen in
order to have the possibility to compare the $I-V$ characteristics
in two perpendicular directions for the same magnetic landscape.

    \begin{figure}[t!]
    \includegraphics[width=80mm]{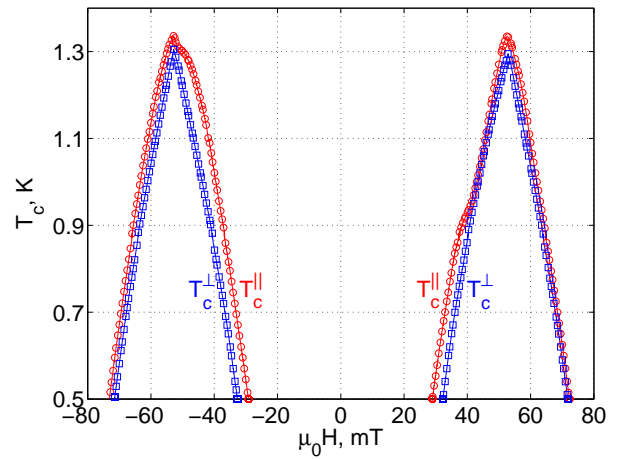}
    \caption{(color online) The phase transition lines $T_c(H)$
    estimated according to the criterium $V(H,T_c)/I_0=0.8\,R_n$ for
    the measurements of magnetotransport using the contacts I and II
    (along the domain wall, red circles) and the contacts III and IV
    (across the domain walls, blue squares). $I_0=100~\mu$A is the dc
    bias current and $R_n$ is the normal state resistance. }
    \label{Fig-PB}
    \end{figure}

    \begin{figure*}[t!]
    \includegraphics[height=60mm]{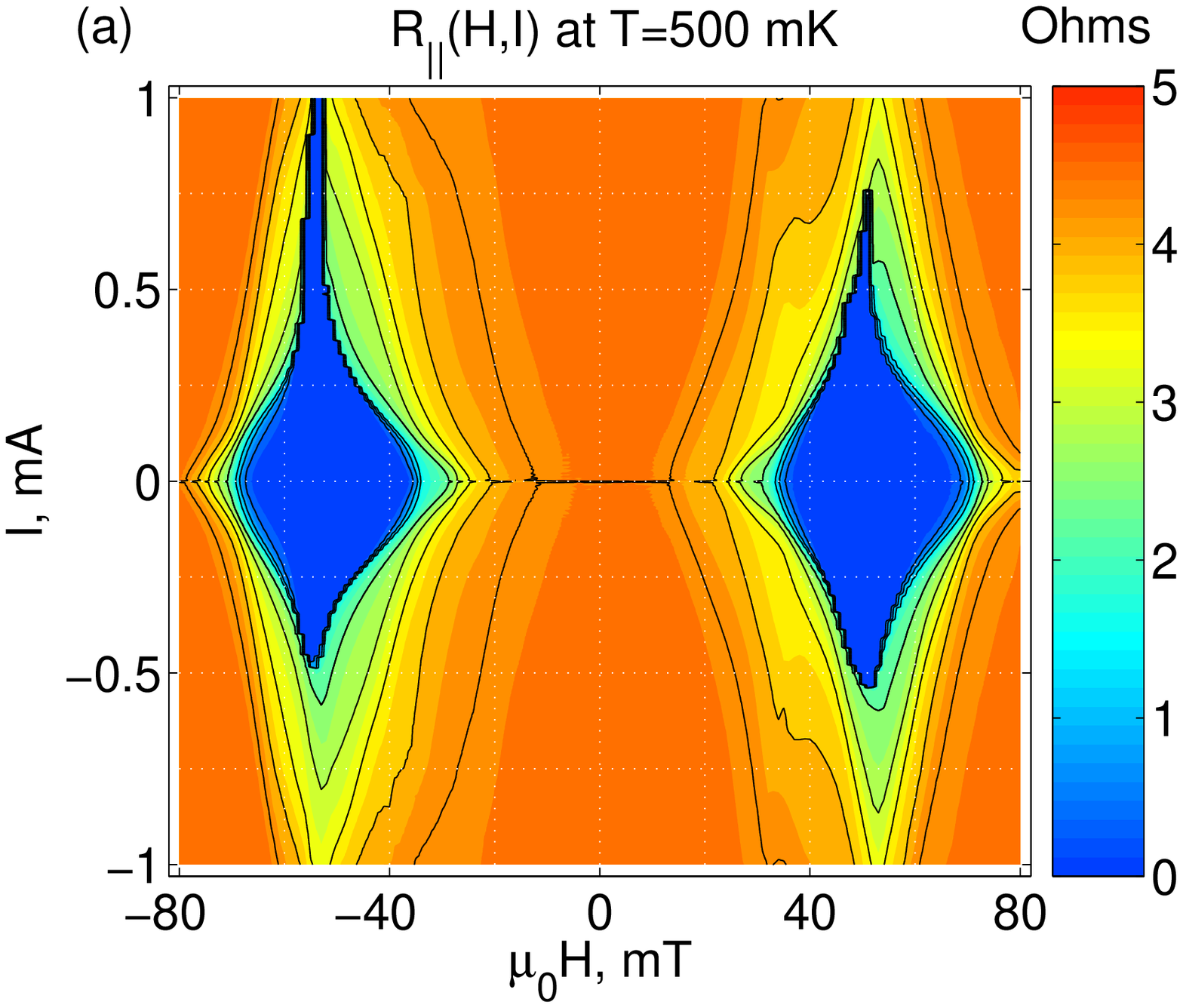}
    \includegraphics[height=60mm]{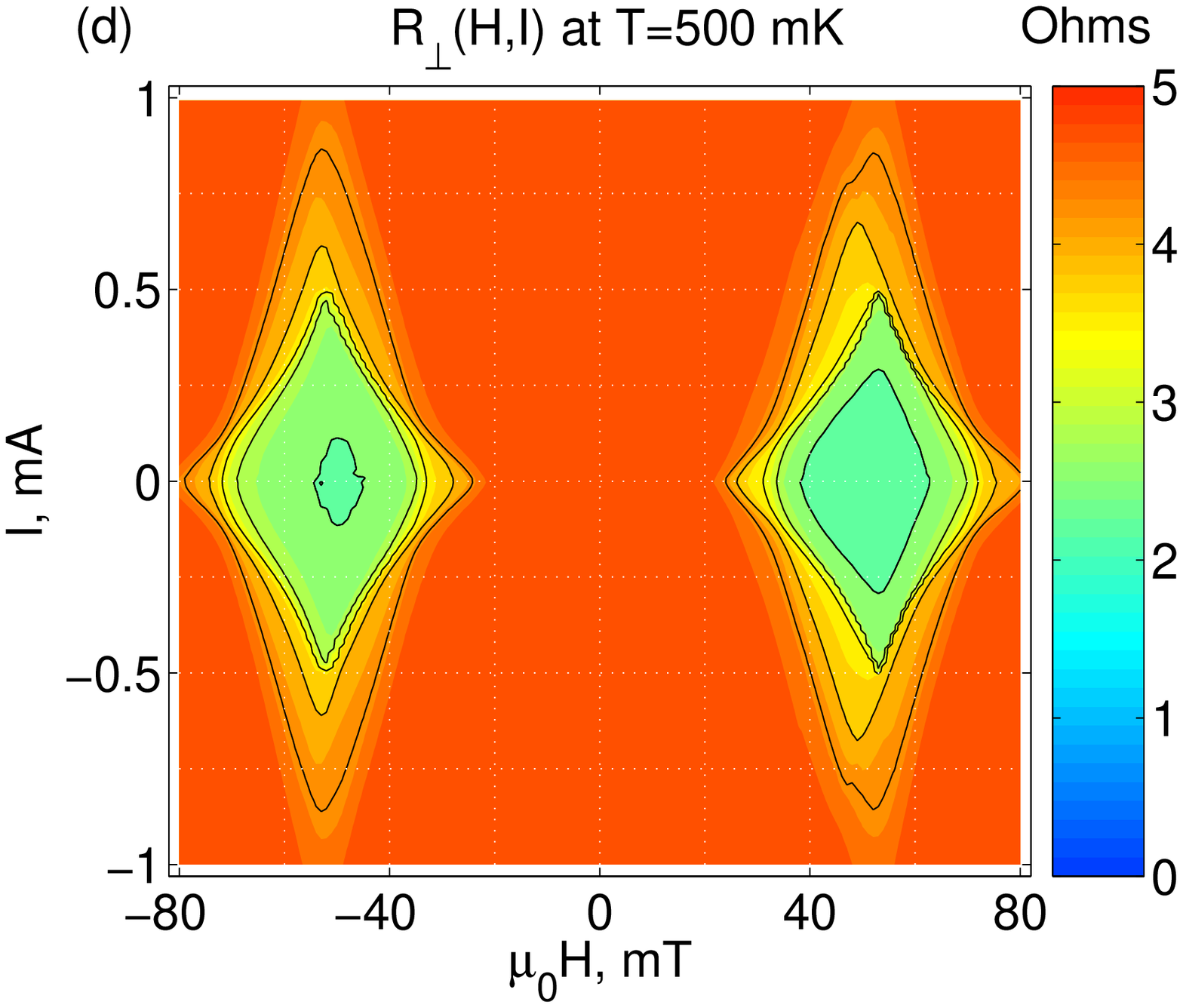}
    \includegraphics[height=60mm]{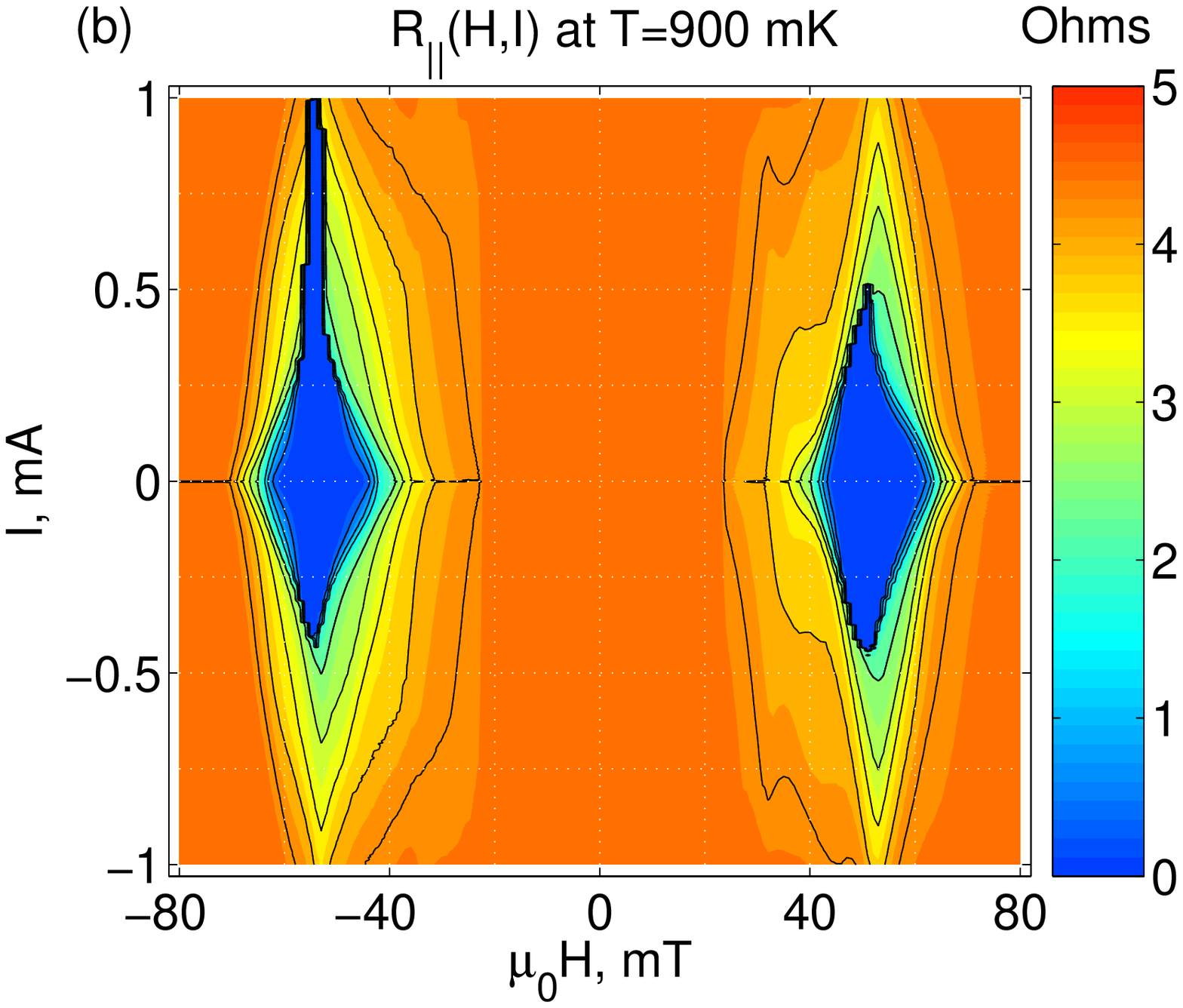}
    \includegraphics[height=60mm]{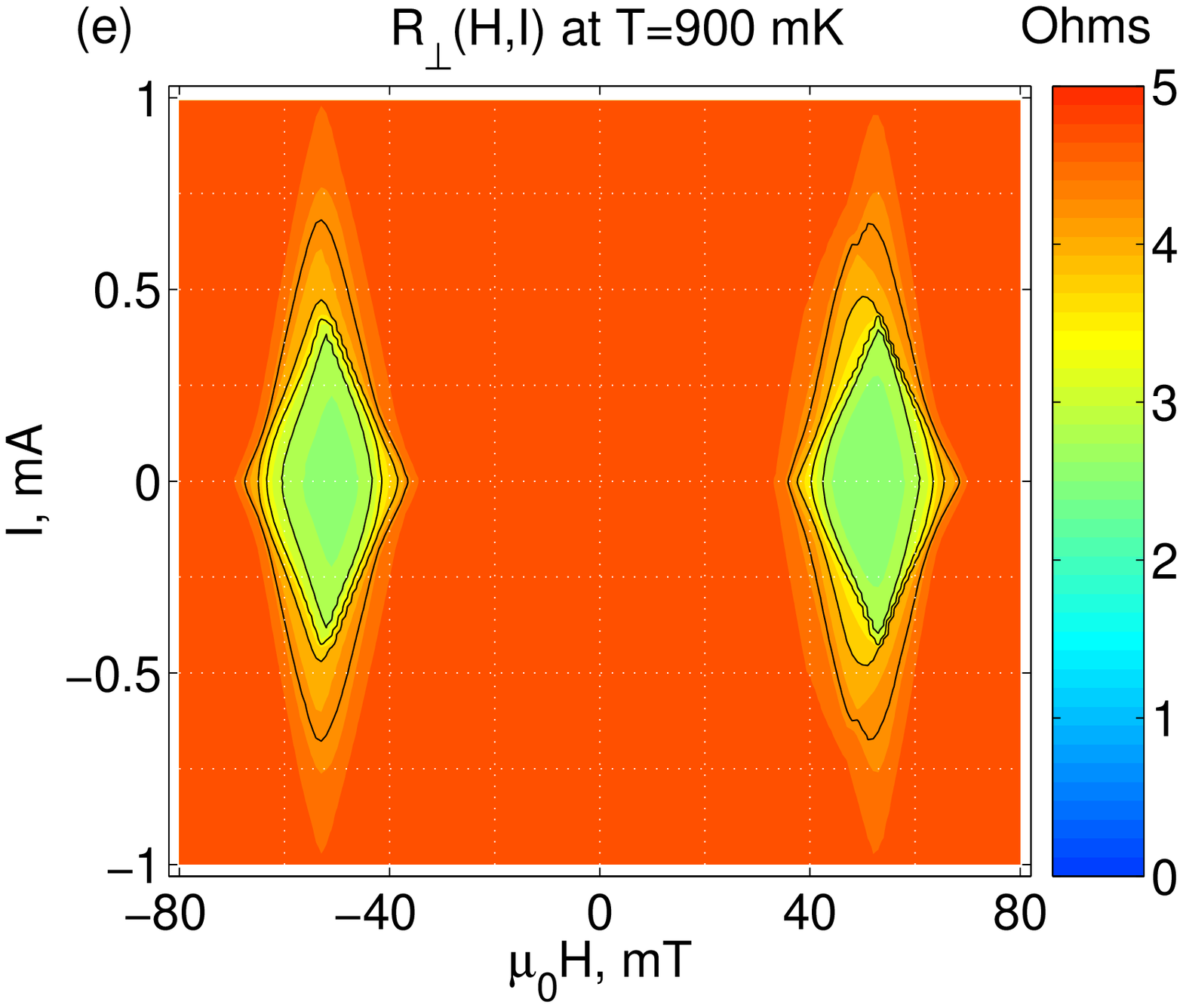}
    \includegraphics[height=60mm]{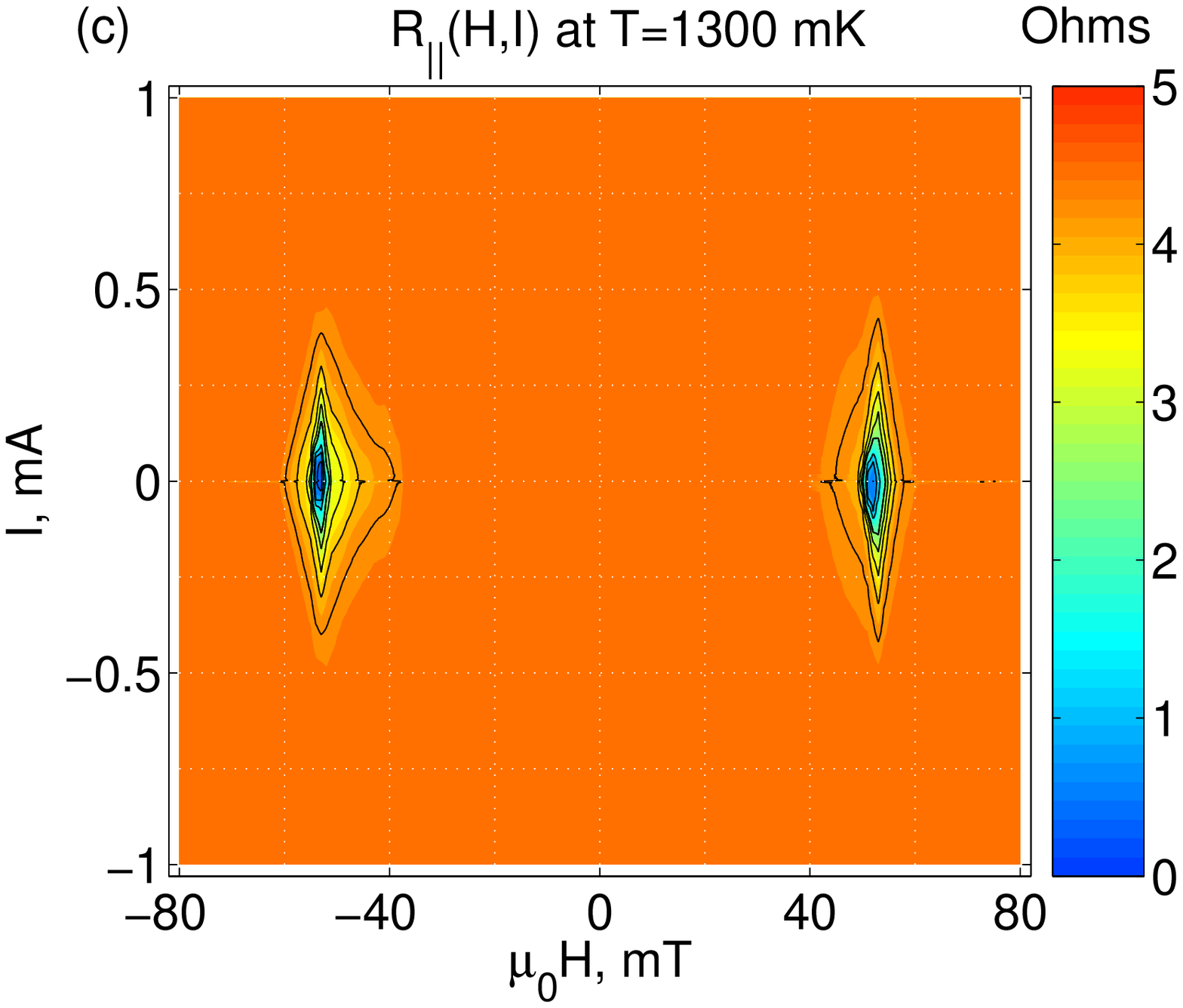}
    \includegraphics[height=60mm]{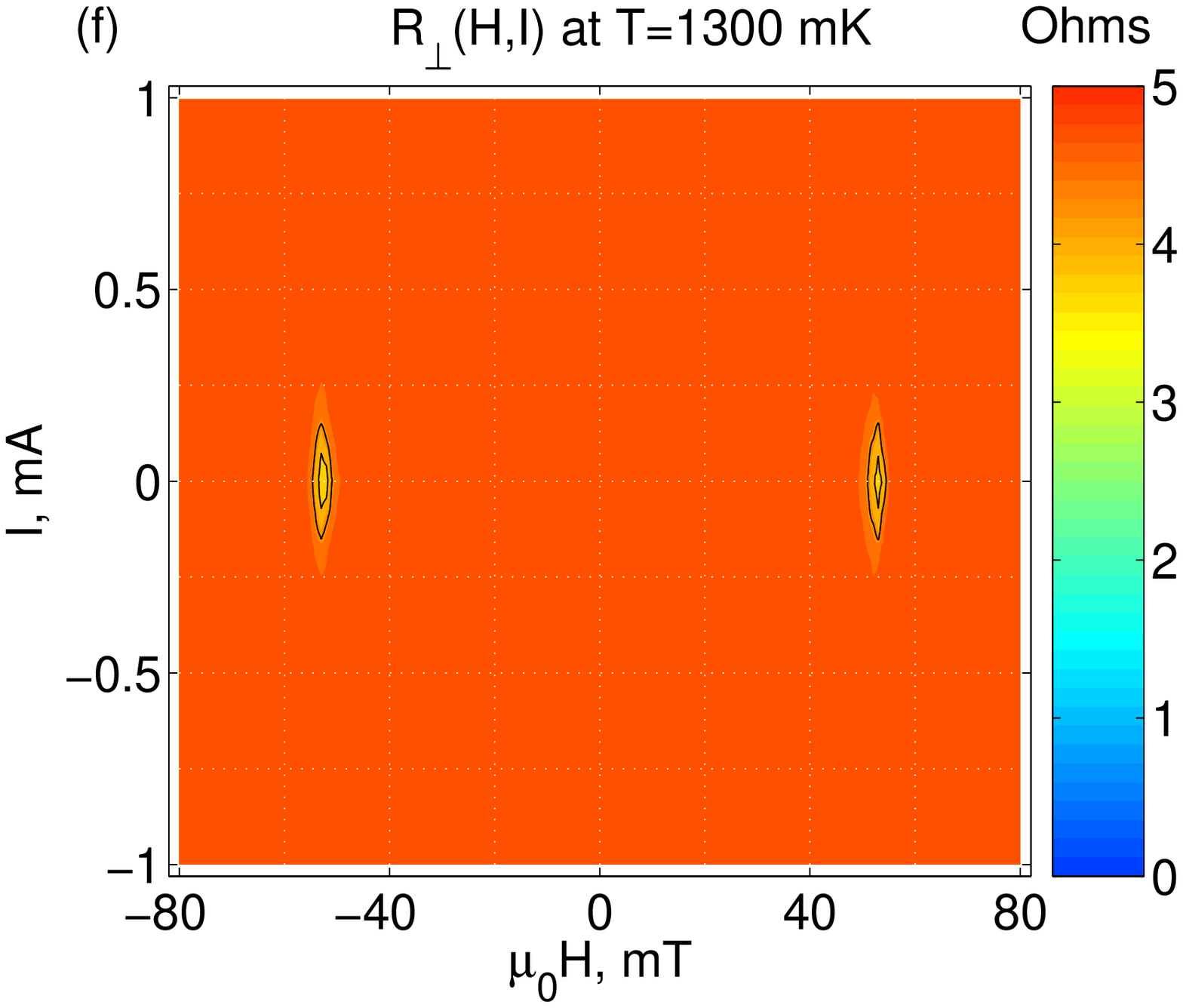}
    \caption{(Color online) The dc resistance $R$ of the
    superconducting bridge as a function of the external magnetic
    field $H$ and the biased dc current $I$, measured along the domain
    wall [panels (a)--(c)] and across the domain
    wall [panels (d)--(f)]. (a) and (d) $T=500$
    mK, (b) and (e) $T=900$ mK, (c) and (f) $T=1300$ mK. Solid black lines are the curves of constant resistance:
    $R(H,I)=0.5$, \,1.0, \,1.5, \,2.0, \,2.5, \,3.0, \,3.5, \,4.0 and 4.5 Ohm. Note that the
    color scales of all plots are identical.     } \label{Fig-Experiment}
    \end{figure*}

    \begin{figure}[t!]
    \includegraphics[width=80mm]{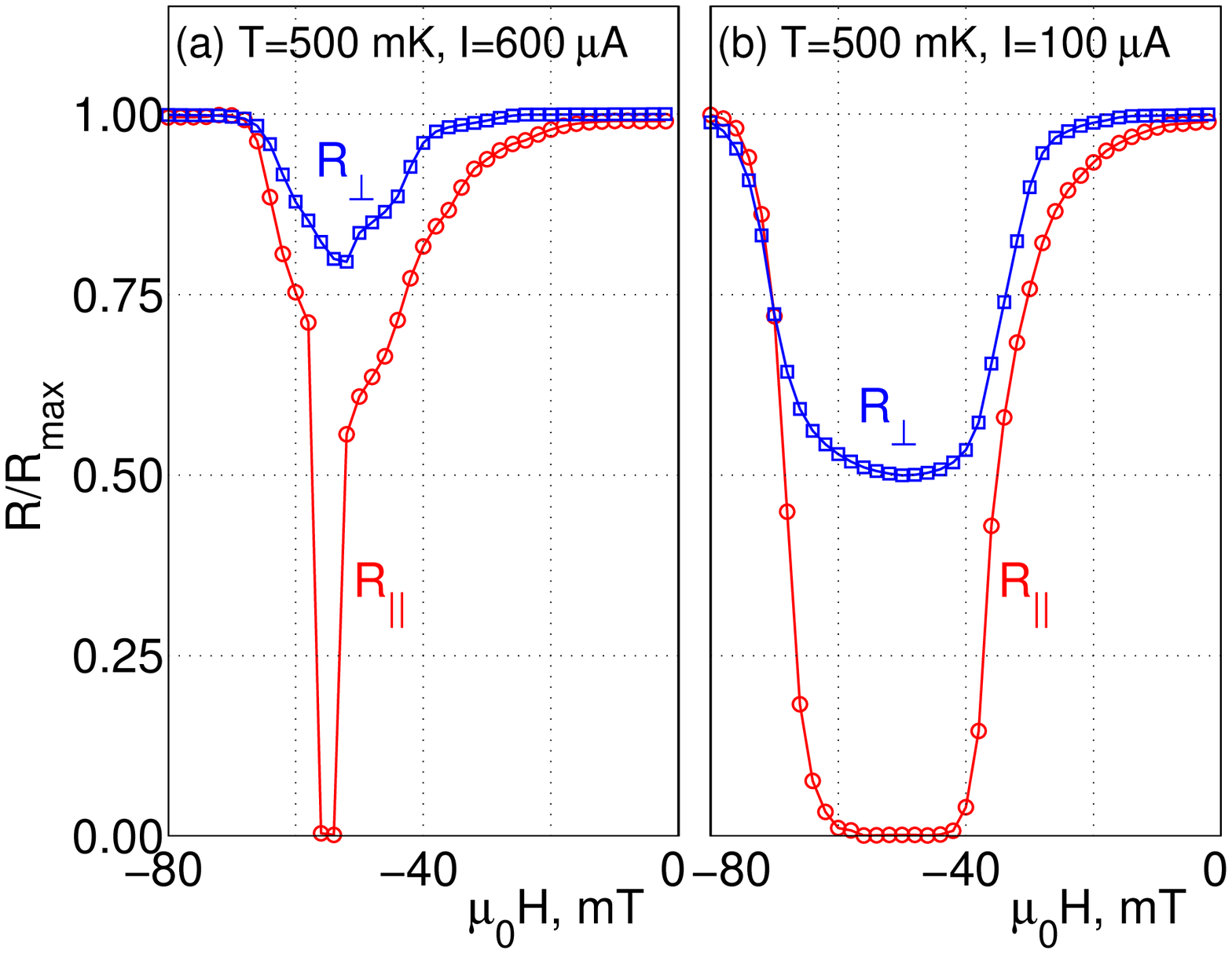}
    \caption{(Color online) Giant resistance anisotropy
    illustrated by cross-sections of $R_{||}(H,I)$ [panel (a) in Fig. \ref{Fig-Experiment}] and
    $R_{\perp}(H,I)$  [panel (d) in Fig. \ref{Fig-Experiment}] taken at $T=500$ mK
    and $I=600 \mu$A (a) and $I=100 \mu$A (b).
    Red circles (blue squares) correspond to the resistivity along (across) the domain wall.}
    \label{Fig-Cross-sections}
    \end{figure}

Figure~\ref{Fig-PB} shows the level curves of the dc resistance of
the sample, $V_{\parallel}(H,T)/I_0=0.8\,R_n$ and
$V_{\perp}(H,T)/I_0=0.8\,R_n$, $R_n$ being the normal-state
resistance, $I_0=100~\mu$A. These lines can be commonly
interpreted as the dependence of the superconducting critical
temperature $T_c$ on $H$. In spite of some inessential
differences, both phase transition lines $T^{\parallel}_c(H)$ and
$T^{\perp}_c(H)$ have symmetrical maxima of similar amplitudes,
and they are characterized by almost the same slope $dT_c/dH$. In
our opinion, this indicates that the nucleation of
superconductivity, responsible for an initial deviation of the
electrical resistance from its normal value, is almost isotropic
(i.e. independent on the direction in which the bias current was
injected and the voltage drop was recorded). Taking the position
of the $T_c$ maxima and comparing the slope $dT_c/dH$ with that
for the regime of surface superconductivity $dT_{c3}/dH\simeq
0.6\,T_{c0}/H_{c2}^{(0)}$, we estimate the amplitude of the
nonuniform field $B_0\simeq 52$~mT, the upper critical field
$\mu_0 H_{c2}^{(0)}\simeq 20.4$~mT at $T=0$, and the maximal
critical temperature $T_{c0}\simeq 1.35$~K. These values appear to
be typical for pure Al films and bridges \cite{Gillijns-PRB-07}.

However deeper in the superconducting state in the $H-T$ plane the
transport properties of the S/F hybrid system become essentially
anisotropic. Figure~\ref{Fig-Experiment} illustrates this, showing
the dependencies of the resistance $R_{\parallel}=V_{\parallel}/I$
(top row) and $R_{\perp}=V_{\perp}/I$ (bottom row) as a function
of $H$ and $I$, derived from the isothermal $I-V$ curves at
constant $H$ value. As expected the total resistance of the sample
goes to zero only for the parallel geometry when $I$ flows along
domain walls [Fig.~\ref{Fig-Experiment} (a)--(c)]. Indeed, the
stripe-type domain structure allows to form a continuous path for
the supercurrents at $|H|\simeq B_0$, connecting the electrodes I
and II. It is easy to see that the maximal critical current
corresponds to the most effective compensation, when one part of
the bridge is subjected to zero local magnetic field,
$B_z=\mu_0H+b_z\simeq 0$, while $B_z\simeq 2B_0$ induces the
normal state in the other part. Taking $I_{max}=1035~\mu$A and the
sample's cross-section $S=1.5\cdot10^{-8}~$cm$^2$, one can
estimate the critical current density $j_c=2I_{max}/S\simeq
1.4\cdot10^5$~A/cm$^2$ at $T=500$~mK, which can be interpreted as
the depinning current density. Apparently, an increase of
temperature reduces the size of the area of zero resistance in the
$H-I$ plane. By contrast, alternating superconducting and normal
(N) regions, induced by the magnetic template, act as a series of
resistors if $I$ is injected perpendicular to the S-N interfaces.
Consequentially, the formation of superconductivity above the
reverse domains roughly halves the total resistance of the sample
at low temperatures: ${\rm min\,}V_{\perp}/I\simeq R_n/2$
[Fig.~\ref{Fig-Experiment} (d)--(f)]. Therefore, in the vicinity
of the compensation fields, we observe a giant anisotropy of
resistivity: ${\rm min\,}R_{\perp}/{\rm min\,}R_{\parallel}>10^3$
(Fig.~\ref{Fig-Cross-sections}), which is in agreement with that
obtained for a network of parallel magnetic domains in permalloy
films
\cite{Vlasko-Vlasov-PRB-08a,Belkin-APL-08,Belkin-PRB-08,Vlasko-Vlasov-PRB-08b}.

In summary, we demonstrated that a 1D domain structure in a
ferromagnetic substrate can induce a giant anisotropy of the
electrical transport of S films that are placed on top of the
substrate. This effect is caused by the appearance of
superconducting channels that run along the underlying magnetic
domains. We also studied the $H$- and $T$ dependence of the
critical current through such an individual channel.

The authors are grateful to A.S. Mel'nikov, A.A. Fraerman and R.
Kramer for stimulating discussions. This work was supported by the
Methusalem Funding of the Flemish Goverment, NES -- ESF program,
the Belgian IAP, the Fund for Scientific Research -- Flanders
(F.W.O.--Vlaanderen), the Russian Fund for Basic Research, RAS
under the program "Quantum physics of condensed matter" and the
Presidential grant MK-4880.2008.2.

\end{document}